\documentclass[aps,prl,twocolumn,superscriptaddress,amsfonts,amsmath,amssymb,showpacs,floatfix,amsart]{revtex4-2}
\usepackage[pdftex]{graphicx} 
\usepackage{hyperref,dcolumn,longtable}
\usepackage{color}
\usepackage[dvipsnames,svgnames,x11names,hyperref]{xcolor}
\hypersetup{colorlinks,citecolor=blue,filecolor=blue,linkcolor=blue,urlcolor=blue}
\usepackage{float}
\usepackage{footnote}
\usepackage{multirow}
\usepackage{ulem}

\begin{document}

\title{Direct Determination of Fission-Barrier Heights Using Light-Ion Transfer in Inverse Kinematics}

\author{S.~A.~Bennett}
\author{K.~Garrett}
\author{D.~K.~Sharp}
\email[Correspondence to: ]{david.sharp@manchester.ac.uk}
\affiliation{Department of Physics and Astronomy, University of Manchester, Manchester M13 9PL, United Kingdom}
\author{S.~J.~Freeman}
\affiliation{Department of Physics and Astronomy, University of Manchester, Manchester M13 9PL, United Kingdom}
\affiliation{CERN, CH-1211 Geneva 23, Switzerland}
\author{A.~G.~Smith}
\author{T.~J.~Wright}
\affiliation{Department of Physics and Astronomy, University of Manchester, Manchester M13 9PL, United Kingdom}
\author{B.~P.~Kay} 
\affiliation{Physics Division, Argonne National Laboratory, Lemont, Illinois 60439, USA}

\author{T.~L.~Tang}
\altaffiliation{Current address: Department of Physics, Florida State University, Tallahassee, Florida 32306, USA}
\affiliation{Physics Division, Argonne National Laboratory, Lemont, Illinois 60439, USA}

\author{I.~A.~Tolstukhin}
\affiliation{Physics Division, Argonne National Laboratory, Lemont, Illinois 60439, USA}

\author{Y.~Ayyad}
\affiliation{IGFAE, Universidade de Santiago de Compostela, E-15782 Santiago de Compostela, Spain }

\author{J.~Chen}
\affiliation{Physics Division, Argonne National Laboratory, Lemont, Illinois 60439, USA}

\author{P.~J.~Davies}
\affiliation{School of Physics, Engineering and Technology, University of York, Heslington, York YO10 5DD, United Kingdom}

\author{A.~Dolan}
\affiliation{Oliver Lodge Laboratory, University of Liverpool, Liverpool L69 7ZE, United Kingdom}

\author{L.~P.~Gaffney}
\affiliation{Oliver Lodge Laboratory, University of Liverpool, Liverpool L69 7ZE, United Kingdom}

\author{A.~Heinz}
\affiliation{Chalmers University of Technology, SE-41296 G{\"o}teborg, Sweden}

\author{C.~R.~Hoffman}
\affiliation{Physics Division, Argonne National Laboratory, Lemont, Illinois 60439, USA}

\author{C.~M\"{u}ller-Gatermann}
\affiliation{Physics Division, Argonne National Laboratory, Lemont, Illinois 60439, USA}

\author{R.~D.~Page}
\affiliation{Oliver Lodge Laboratory, University of Liverpool, Liverpool L69 7ZE, United Kingdom}

\author{G.~L.~Wilson}
\affiliation{Louisiana State University, Baton Rouge, Louisiana 70803, USA}
\affiliation{Physics Division, Argonne National Laboratory, Lemont, Illinois 60439, USA}

\date{\today}
             
\begin{abstract}
We demonstrate a new technique for obtaining fission data for nuclei away from $\beta$-stability. These types of data are pertinent to the astrophysical \textit{r-}process, crucial to a complete understanding of the origin of the heavy elements, and for developing a predictive model of fission. These data are also important considerations for terrestrial applications related to power generation and safeguarding. Experimentally, such data are scarce due to the difficulties in producing the actinide targets of interest. The solenoidal-spectrometer technique, commonly used to study nucleon-transfer reactions in inverse kinematics, has been applied to the case of transfer-induced fission as a means to deduce the fission-barrier height, among other variables. The fission-barrier height of $^{239}$U has been determined via the $^{238}$U($d$,$pf$) reaction in inverse kinematics, the results of which are consistent with existing neutron-induced fission data indicating the validity of the technique. 
\end{abstract}

\maketitle

The majority of heavy nuclei, including those important in both  terrestrial and astrophysical settings, have no available nuclear data from neutron-induced fission~\cite{schmidt2018review}. For example, the astrophysical \textit{r-}process is thought to account for the creation of approximately half of the heavy elements beyond iron. Alongside other ingredients, fission data such as barrier height and mass and charge yields are crucial inputs in performing accurate abundance calculations for high-$Z$ nuclides~\cite{giuliani2020fission, mumpower, Goriely}. In particular, the so-called fission recycling mechanism defines the upper mass limit in the \textit{r-}process, and channels mass to lower regions of the nuclear chart, thus contributing to the abundances of medium mass nuclei. Moreover, in the era of multi-messenger astronomy, fission data are critical for a deeper understanding of results which suggest that fission is significant in the process of nucleosynthesis in neutron-star mergers \cite{Goriely}. For example, the fission product $^{90}$Sr has recently been observed in the remnants of such an event~\cite{watson2019identification}.

Current benchmarking of fission models is performed indirectly, for example by comparing the limits of the neutron capture process in nuclear explosions to those predicted in calculations \cite{mollercalc}. Validating these models for the case of nuclei with short half-lives is not feasible due to the impracticability of producing fixed targets. For direct validation, alternative methods are required, for example by using the fissioning system as a beam to obtain experimental data. Furthermore the benchmarking of fission models \cite{mollercalc1,mollercalc,Erlercalc} and collection of fission data for short-lived actinides, including fission barriers and mass and charge yields, are key quantities for future terrestrial power and safeguarding applications \cite{kolos2022current,farget2015transfer}.
\begin{figure*}
\includegraphics[width=13.0cm]{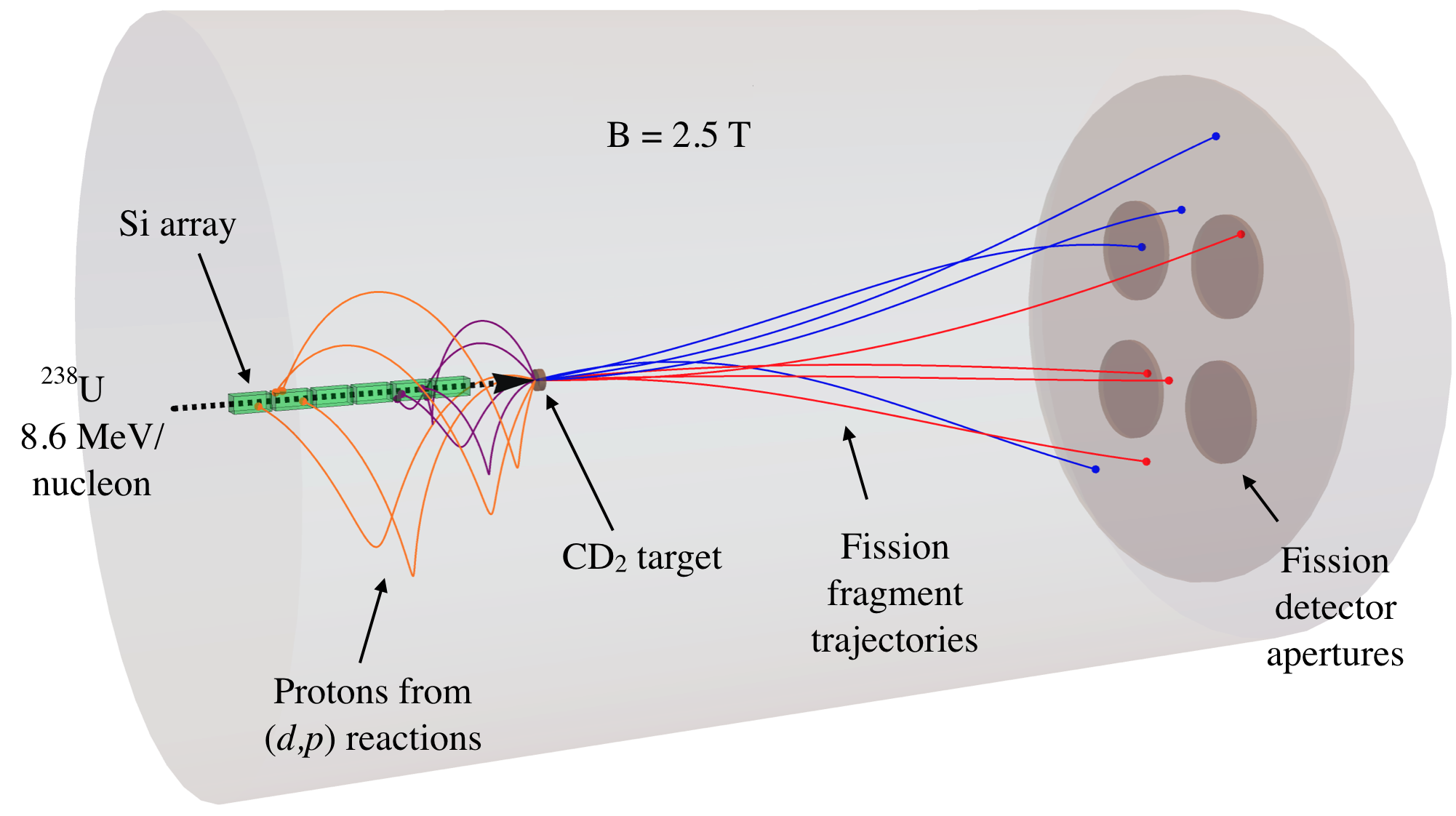}
\caption{\label{fig:setup} To-scale schematic of the experimental setup with example particle trajectories for $^{238}$U($d$,$pf$) events. Example proton trajectories for reactions populating the ground state in $^{239}$U (orange curves) and states at 7~MeV close to the fission barrier  (purple curves) are shown for a range of c.m. proton angles. Example fission fragment trajectories are also shown for fragments with $A = 138$ (red curves) and $A=100$ (blue curves), for a range of emission angles. The equally spaced circular detector apertures have radius 8~cm, and are centered 18~cm from the beam axis. The axial distance between the target and detector apertures is 70~cm.}
\end{figure*}

Any reaction where the $Q$ value can be accurately determined and that produces the excited compound system of interest can be used to probe the fission barrier. Direct reactions can, for example, be used to populate single-particle doorway states leading to compound nucleus formation. In the ($d$,$p$) reaction, the neutron transferred from the deuteron to the target nucleus acts as a proxy for the neutron-induced reaction. By studying the ($d$,$pf$) reaction, information can be obtained about the fission-barrier height. The probability of fission and the neutron-induced fission cross section can also be deduced, assuming that the fission cross section $\sigma_{{f}}^{A}$ can be factorized into a compound nucleus formation cross section $\sigma_{\mathrm{CN}}^{A+1}$ and fission decay probability $P_{{f}}^{A+1}$:
\begin{equation}\label{eqn:fissxs}
\sigma_{{f}}^{A} = \sigma_{\mathrm{CN}}^{A+1} \times  P_{{f}}^{A+1}.
\end{equation} 
The technique was first demonstrated in normal kinematics by Northrop \textit{et al.}~\cite{northrop1959measurement} and by Cramer and Britt~\cite{cramer1970fission}, where the fixed target was the actinide species of interest.

There are numerous examples of reaction-induced fission experiments from which the fission-barrier height, fission probability, and other properties such as mass split have been accurately deduced. In normal kinematics, examples include: ($d$,$pf$)~\cite{ducasse2016investigation,britt1970fission}, ($t$,$pf$)~\cite{cramer1970fission, britt1970fission}, ($^{3}$He,$\alpha f$)~\cite{kessedjian2010neutron}, and in inverse kinematics: ($^{9}$Be,$^{8}$Be~$f$)~\cite{ramos2019first,rodriguez} and multinucleon transfer~\cite{rodriguez,leguillon2016fission,caamano2013isotopic}. A review of the technique can be found in Ref.~\cite{escher2012compound}, where questions around the degree to which direct reactions can act as true ``surrogates" for neutron-induced compound reactions are also discussed. For example, in near-barrier fission, the low level density leads to a strong sensitivity of the fission probability to the spin-parity distribution of states populated in the transfer reaction, which are plausibly very different to those of neutron-induced reactions \cite{escherratio,chiba2010verification}. In general however, results compare remarkably well with neutron-induced data.

In this work, we present a study of the transfer-induced fission of $^{239}$U in inverse kinematics with the ($d$,$pf$) reaction using the solenoidal-spectrometer technique. This is presented as an an exploratory case where there are existing fission data, and constitutes the first direct measurement of a fission-barrier height using a light-ion transfer reaction in inverse kinematics. Experiments performed in such a way, using unstable species as a beam, clearly permit studies of a large number of nuclei that are not accessible in fixed-target experiments in normal kinematics, nor by using multinucleon transfer on stable beams in inverse kinematics. There exist three solenoidal spectrometers used for transfer reaction studies: HELIOS at Argonne National Laboratory (ANL)~\cite{lighthall2010commissioning}, the ISOLDE Solenoidal Spectrometer at CERN~\cite{macgregor2021evolution}, and SOLARIS at the Facility for Rare Isotope Beams (FRIB)~\cite{solaris}. There are many available radioactive beams of sufficient intensity at the facilities hosting such devices to which the method presented in this paper may be applied. In the longer term, it is envisaged that more exotic, neutron-rich isotopes will become available with developments at these facilities, for example the reaccelerator (ReA) at FRIB~\cite{frib}, and the Laser Ionization and Spectroscopy of Actinides (LISA) project in Europe (ISOLDE)~\cite{lisa}.

The experiment was carried out using HELIOS at ANL. The geometry is shown in Fig.~\ref{fig:setup}, along with the calculated trajectories of reaction products. Light ions emitted following reactions of the beam with a deuterated polyethylene (CD$_2$) target are transported to a four-sided position-sensitive silicon array situated upstream of the target surrounding the beam axis, which is itself collinear with a solenoidal magnetic field. The Si array is used to measure the laboratory energy of light ejectiles and their return distance to the beam axis. The 350-mm long Si array was positioned such that the end was 55~mm upstream of the target. In this configuration protons from ($d$,$p$) reactions populating states at an excitation energy of 7~MeV emitted at angles from approximately 10$^{\circ}$ to 30$^{\circ}$ in the center-of-mass  frame (c.m.) hit the array. The Si array was calibrated in energy using $\alpha$ particles from a $^{228}$Th source. Downstream of the target, a Faraday cup (not shown in Fig.~\ref{fig:setup}) was used to measure beam current, and an annular silicon detector was used to detect elastically scattered deuterons for a narrow range of c.m. angles between 29$^{\circ}$ and 29.3$^{\circ}$. The latter allows an absolute normalization of measured yields to generate absolute cross sections.

A set of gas-filled heavy-ion detectors was used downstream of the target to study the subset of residual $^{239}$U nuclei that fission. In inverse kinematics, the fission-fragment angular distribution is strongly forward peaked in the laboratory as indicated by the example trajectories shown in Fig.~\ref{fig:setup}. With a beam energy of 8.6~MeV/nucleon, light and heavy fragments from the fission of actinide nuclei form two cones at laboratory angles of approximately 15$^{\circ}$ and 10$^{\circ}$ respectively. An aluminium charge-reset foil with a thickness of 100~$\mu$g/cm$^2$ was positioned 70~mm downstream of the target in order to minimise the spread of charge states of fission fragments emitted at different depths inside the CD$_2$ target. Four fission detection arms were positioned around 1~m downstream of the target, two at 15$^{\circ}$ to the beam axis and two at 10$^{\circ}$ to maximize the acceptance of light and heavy fragments. Each of the arms had a position-sensitive multi-wire proportional counter (MWPC) followed by a gaseous axially-segmented Bragg detector. On an event-by-event basis, the Bragg detector yielded the total and specific energy-loss of any detected fragments, and the MWPCs gave their position at the entrance to the Bragg detectors. The position was internally calibrated using the $\alpha$ particles from the $^{228}$Th source. The MWPCs were also used to generate a precise timing signal used to correlate fission fragments to Si array events. The intrinsic efficiency was close to 100\%. The geometric efficiency was $\sim$10\% for the detection of one or more fission fragments, and $\sim$1\% for the coincident detection of light and heavy fragments (more details below). The energy loss for typical light on the efficiency considerations for the fission fragment detectors. On an event-by-event basis, the information yielded by the fission detectors is sufficient to deduce the fission mass split and c.m. fission axis orientation, through kinematic reconstruction, and the atomic number of each fragment using Bragg-peak spectroscopy~\cite{gruhn1982bragg}. These capabilities allow for the simultaneous measurement of fission yields with excitation energy and fission probability, and will be reported in a future publication.

Placing the Si array upstream of the target and fission fragment detectors downstream leads to precise selection of transfer-induced fission events. With this arrangement, background events such as fission preceded by the evaporation of light charged particles from compound systems is suppressed. To travel into the backwards hemisphere, evaporated particles must possess a large c.m. energy at least as large as the c.m. energy of the beam, around $10$~MeV/nucleon here.

A beam of $^{238}$U$^{47+}$ at an energy of 2.05 GeV (8.6~MeV/nucleon) was delivered by the ATLAS accelerator. The average beam intensity was $\sim10^6$ pps, with a total integrated beam dose of around 5.5~$\times 10^{11}$ ions. In this time, around 3.5~$\times 10^{5}$ ($d$,$p$) events were detected with the Si array, and around 1000 ($d$,$pf$) events were recorded. Due to target damage from the heavy beam, 11 CD$_2$ targets were used during the experiment with thicknesses between 410 and 590~$\mu$g/cm$^2$. Data were also collected with a pure $^{\mathrm{nat}}$C target (thickness 584~$\mu$g/cm$^2$) to evaluate backgrounds arising from multinucleon transfer-induced fission reactions on carbon in the CD$_2$ target~\cite{caamano2013isotopic}. A 2.5-T magnetic field was used, and the digital data acquisition was triggered either by signals in the Si array or any of the MWPCs.
\begin{figure}[h]
\includegraphics[width=8.6cm]{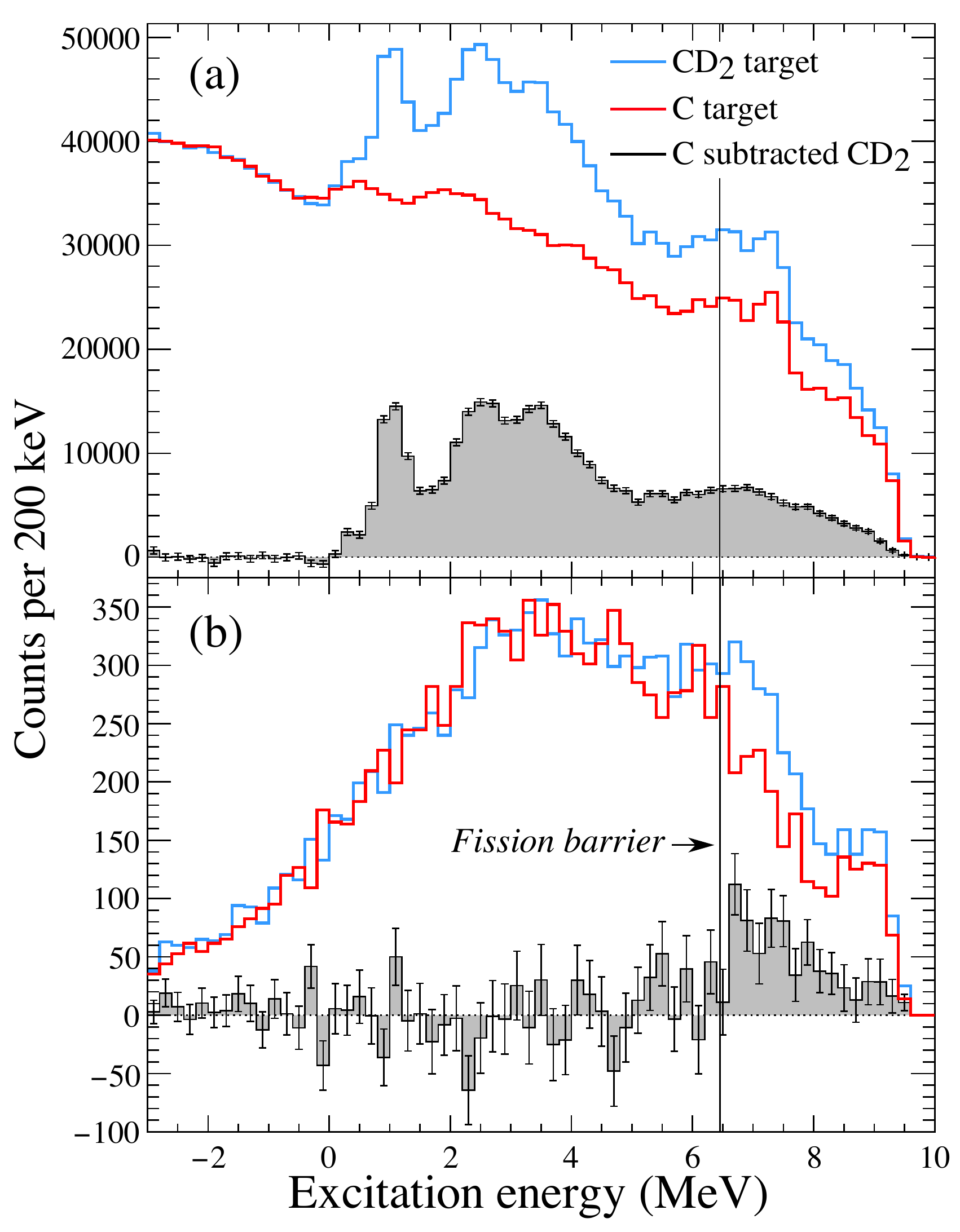}
\caption{(a)~Excitation-energy spectra associated with all events for Si array data taken with both CD$_2$ and C targets where the C data has been scaled onto the CD$_2$ data, and for the carbon subtracted CD$_2$ data. (b)~Same as for (a), but for events in which $\geq 1$ fission fragments are detected with the MWPCs of the fission array. The vertical solid line denotes the known fission-barrier height \cite{bjornholm1980double}. The carbon target spectra were scaled by $\times$1.92 and $\times$1.76 for the singles and fission-gated data, respectively.}
\label{fig:counts1}
\end{figure}

$Q$-value spectra were generated with the measurements of the charged particles detected in the Si array. The $Q$ value was deduced by calculating the c.m. energy of the light ejectile, which itself depends linearly on the laboratory energy and axial return distance to the beam axis \cite{wuosmaa2007solenoidal}. Coincidence events between the Si array and fission detectors were generated by applying a 250-ns wide gate on array-MWPC timing signals. $Q$-value spectra are shown in Fig.~\ref{fig:counts1}. The position of the Si array and a lower energy threshold of 0.5~MeV restricted the measurement to residual nuclei with excitation energy less than around 9.6~MeV. The $Q$-value resolution was around 250~keV (FWHM). An unresolved multiplet of excited states in $^{239}$U around 1~MeV is visible, beyond which there is a continuum. The carbon content of the CD$_2$ target led to a significant background in both the singles and fission-gated spectra. It is likely that this background is due to multinucleon transfer-induced fission reactions where light ejectiles intercept the Si array. This background was addressed by evaluating its shape using the data taken with the pure C target. The resulting carbon spectra were scaled to match the spectra obtained with the CD$_2$ target in the region below the ground state, and subtracted as shown in Fig.~\ref{fig:counts1}. Although it might be expected that the carbon target scaling factor for array singles and fission-gated data are similar, the scaling was done separately for each case as the respective factors were found to be different. This suggests the presence of an additional contribution to the array singles data beyond ($d$,$p$) reactions, most likely deuteron breakup (see discussion below). For array singles, the data were scaled by normalizing over the region $-3\leq E_{x} \leq -0.5$~MeV, and for the fission-gated data between $-3\leq E_{x} \leq 2$~MeV. The background of time-random Si~array-MWPC coincidences was constant with excitation energy, and constitutes 0.43\% of the coincident events within the 250-ns time window.

The fission-barrier height was deduced by constructing the fission probability as a function of excitation energy. The probability $P_f$ for the residual nucleus of excitation energy $E_x$ to decay via fission was determined by the ratio
\begin{equation}\label{eqn:pf}
P_f(E_x) = \frac{N_{d,pf}(E_x)}{N_{d,p}(E_x) \cdot \epsilon_f}
\end{equation} 
where $N_{d,pf}(E_x)$ is the number of ($d$,$p$) events in coincidence with the detection $\geq$1 fission fragments and $N_{d,p}(E_x)$ is the total number of ($d$,$p$) events. The efficiency for the detection of protons from ($d$,$p$) events cancels in the ratio, but the fission detection efficiency $\epsilon_f$ must be taken into account for a proper normalization. This was derived from a simulation accounting for the reaction kinematics, geometry, ion transport in the solenoidal magnetic field, and average charge state distributions of the fission fragments. Fission-fragment $A$ and $Z$ yields and average kinetic energy distributions were taken from GEF~\cite{schmidt2016general}, and the fission axis was assumed to be oriented isotropically in the c.m. frame.

\begin{figure}[]
\includegraphics[width=8.6cm]{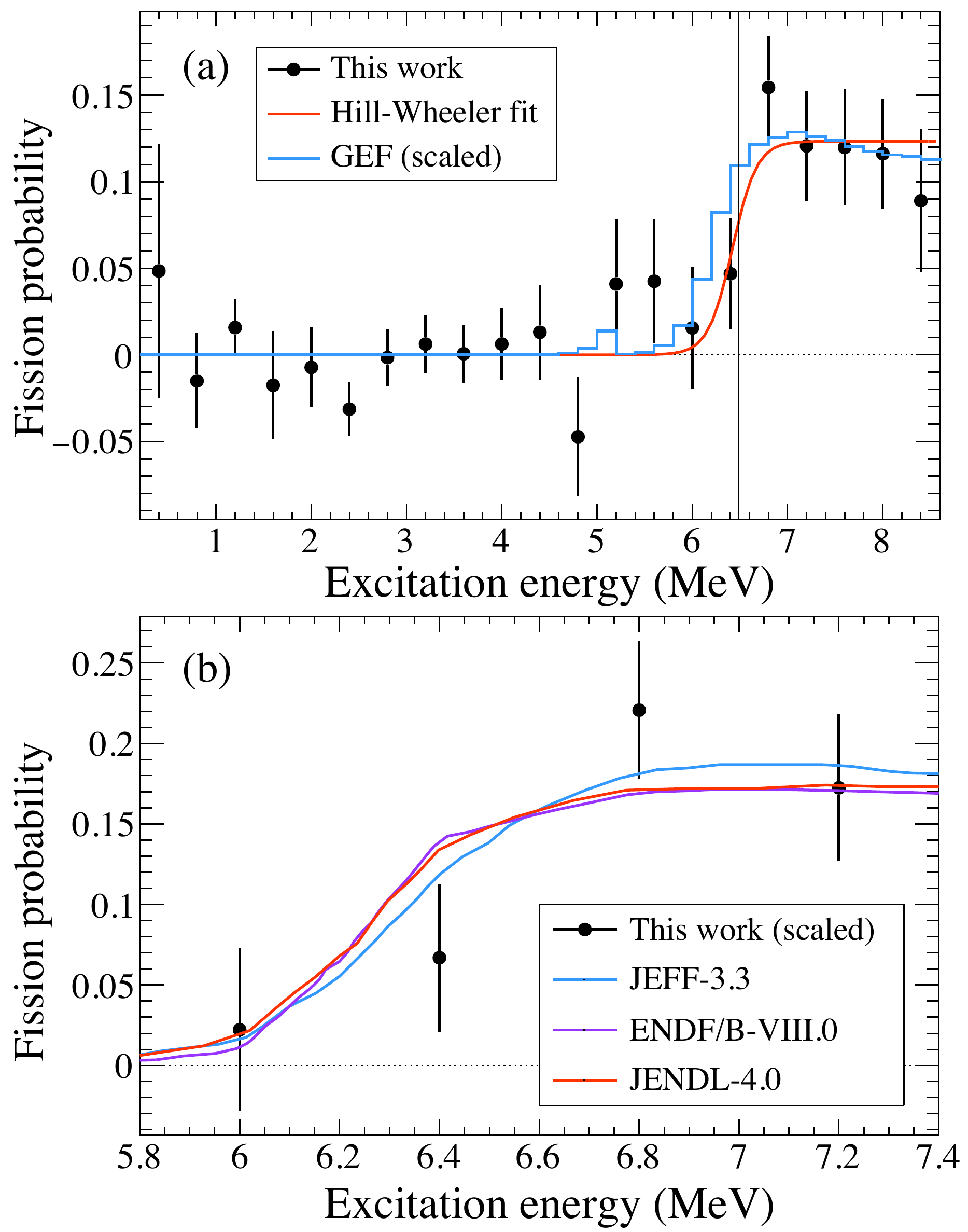}
\caption{(a)~Experimental fission probability, as defined in Equation~\ref{eqn:pf} compared to a GEF simulation~\cite{schmidt2016general} (the GEF result has been normalized to our data) and empirical Hill-Wheeler fit with $\chi^2_{\mathrm{red.}} = 0.57$. The vertical line denotes the known fission-barrier height. (b)~Experimental fission probability in the region around the fission barrier compared to probabilities deduced from evaluated nuclear data libraries (JEFF-3.3~\cite{plompen2020joint}, ENDF/B-VIII.0~\cite{brown2018endf}, JENDL-4.0~\cite{shibata2011jendl}). In the bottom panel, the experimental data have been increased by 30\% as explained in the main text. The error bars represent the statistical uncertainty.}
\label{fig:prob} 
\end{figure}
The deduced fission probability is shown in Figure~\ref{fig:prob}. The data taken with both target materials were statistically limited, however an increase in the fission probability between 6~and~7~MeV corresponding to the region around the known fission barrier ($B_f=6.46$~MeV~\cite{bjornholm1980double}) is unambiguous and is consistent with the fission barrier deduced from neutron-induced fission data evaluations, also illustrated in Figure~\ref{fig:prob}. This is corroborated with a fit of a Hill-Wheeler function of the form $P_f(E_{x}) = A \times\lbrace 1+\exp(\frac{2\pi}{\hbar \omega}[B_f - E_{x}]) \rbrace ^{-1}$. The parameter $A$ represents the fission probability above the barrier, $\hbar \omega$ is the diffuseness of the barrier and $B_f$ is the height of the fission barrier. For the fit to our data, the diffuseness was fixed taking the value from Ref.~\cite{bjornholm1980double} ($\hbar \omega = 0.8$~MeV), due to a lack of data points defining the region around the barrier. The barrier height in this fitting procedure is, however, largely independent of the diffuseness. The fit gave values of $A=0.123(15)$ and $B_f = 6.42(12)$~MeV, consistent with the value of $B_f = 6.46$~MeV from Ref.~\cite{bjornholm1980double}. The shape of the measured fission probability is consistent with the fission probability from GEF~\cite{schmidt2016general} as well as data evaluations \cite{plompen2020joint,brown2018endf,shibata2011jendl}. The absolute magnitude of the fission probability above the barrier is found to be lower than data evaluations by around 30\%. This effect, at the same order of magnitude, has been observed in similar studies in normal kinematics~\cite{ducasse2016investigation} and is attributed to deuteron breakup. Protons from breakup reactions lead to a surplus of Si array events and, if not accounted for, are interpreted as being ($d$,$p$) events. This leads to an underestimation of the fission probability. The magnitude of this effect cannot be experimentally determined; theoretical calculations are required to obtain correction factors, see for example Ref.~\cite{ducasse2016investigation}. A further effect, on the level of a few percent, is the angular anisotropy of the fission axis in the c.m. frame which in principle affects the fission-fragment detection efficiency $\epsilon_f$, see for example Ref. \cite{pal2017projectile}. Both corrections, deuteron breakup and fission anisotropy, are not required to extract a value for the fission-barrier height reported in this work. The ($d$,$p$) and ($d$,$pf$) cross sections are available as Supplemental Material~\cite{supp}.

In conclusion, a fission-barrier height has for the first time been determined using a light-ion transfer reaction in inverse kinematics, in this work for the case of $^{239}$U using the solenoidal spectrometer technique, thus unambiguously demonstrating the validity of a technique that can be applied to other cases of interest. By performing similar experiments at the radioactive ion beam facilities with solenoidal spectrometers, this technique could be used to address the scarcity of fission data for a range of short-lived nuclei. Such data are required to benchmark and assess the validity of fission models pertinent to the astrophysical \textit{r-}process and, in particular, the so-called fission recycling mechanism. It is envisaged that studies of this nature are therefore of fundamental importance to a complete understanding of the origins of the heavy elements, as well as  being a powerful tool for gathering nuclear data relevant to terrestrial applications. \\

\textit{Acknowledgements}.--- This work was supported by the UK Science and Technology Facilities Council under Grant Numbers ST/P004598/1 and ST/V001027/1 (Liverpool), ST/P004423/1 and ST/V001116/1 (Manchester), ST/R004056/1 (Gaffney), ST/T004797/1 (Sharp), ST/N00244X/1 (UK Nuclear Data Network), by the Knut and Alice Wallenberg Foundation Dnr. KAW 2020.0076, and by the U.S. Department of Energy, Office of Science, Office of Nuclear Physics, under Contract Numbers DE-AC02-06CH11357. This research used resources of ANL’s ATLAS facility, which is a DOE Office of Science User Facility. The authors would like to acknowledge the support of the ATLAS operations team in the development and delivery of the beam. The supporting data are openly available at Ref. \cite{data}.

\end{document}